\documentclass[11pt]{article}
\usepackage{amssymb}

\newcommand{\bi}{\bibitem}
\newcommand{\be}{\begin{eqnarray}}
\newcommand{\ee}{\end{eqnarray}}
\newcommand{\rar}{\rightarrow}

\begin{document}


\begin{titlepage}

\title{Baryogenesis from Gravitational Decay of TeV-Particles
in Theories with Low Scale Gravity}

\author{C. Bambi$^{a,b,}$\footnote{E-mail: bambi@fe.infn.it},
A.D. Dolgov$^{a,b,c,}$\footnote{E-mail: dolgov@fe.infn.it}
and K. Freese$^{d,}$\footnote{E-mail: ktfreese@umich.edu}}

\maketitle

\begin{center}
$^{a}$Istituto Nazionale di Fisica Nucleare, Sezione di Ferrara,
       I-44100 Ferrara, Italy\\
$^{b}$Dipartimento di Fisica, Universit\`a degli Studi di Ferrara,
       I-44100 Ferrara, Italy\\
$^{c}$Institute of Theoretical and Experimental Physics,
       113259, Moscow, Russia\\
$^{d}$Michigan Center for Theoretical Physics, Physics Dept., University
       of Michigan, Ann Arbor, MI 48109
\end{center}

\vspace{0.5cm}

\begin{abstract}
In models with the fundamental gravity scale in the TeV range, early
cosmology is quite different from the standard picture, because the
universe must have arisen at a much lower temperature and the
electroweak symmetry was probably never restored. In this context,
baryogenesis appears to be problematic: if the involved physics is
essentially that of the Standard Model, ``conventional''
non-conserving baryon number processes are completely negligible at
such low temperatures.  In this paper we show that the observed
matter-antimatter asymmetry of the universe may be generated by
gravitational decay of TeV-mass particles: such objects can be out of
equilibrium after inflation and, if their mass is of the same order of
magnitude as the true quantum gravity scale, they can quickly decay
through a black hole intermediate state, violating global symmetries,
in particular, baryon number. In this context, we take advantage of
the fact that the ``Sakharov conditions'' for baryogenesis can be more
easily satisfied with a low fundamental scale of gravity.
  
\end{abstract}

\end{titlepage}


\section{Introduction}

The mechanism of generation of the observed matter-antimatter
asymmetry in the universe is an open problem in modern cosmology and a
clear sign of new physics beyond the Standard Model. Many
possibilities have been proposed~\cite{b-rev}, but at present there is
no experimental evidence in favor of one model over another. In
addition, unluckily, the models are often based on assumptions
difficult to test, since the involved physics is at such high energies
as to be unreachable in future laboratories on the Earth.

We will consider baryogenesis scenarios based on low scale gravity
with the fundamental Planck mass, $M_*$, in the TeV range. As we shall
see in what follows, the baryogenesis scenarios in models with a low
gravity scale encounter some additional problems, because of an
expected very low reheating temperature of the universe, and,
therefore, additional exotic assumptions, e.g. time variation of
fundamental constants, are usually needed.  In this context, on the
other hand, we will show that the ``Sakharov conditions'' for
baryogenesis can be more easily satisfied with a low fundamental scale
of gravity.  In addition, the mechanism may operate with the minimal
particle content (only known quarks) or with a minor extension to low
energy supersymmetry.

In the standard framework of general relativity, there is probably no
realistic possibility to ever observe gravitational interactions in
particle physics. This is due to the fact that there are apparently
two distinct fundamental energy scales in nature which are different
by many orders of magnitude, namely, the Planck mass $M_{Pl} \sim
10^{19}$ GeV, which sets the energy when gravity becomes comparable to
gauge interactions, and the electroweak scale of the Standard Model of
particle physics, $M_{EW} \sim 10^{3}$ GeV, which is accessible in
lepton and hadron colliders.

However, the interpretation of $M_{Pl}$ and $M_{EW}$ as two
fundamental energy scales may be incorrect because the previous
assertion is based on the non-trivial assumption that gravity behavior
is unchanged down to the Planck length $L_{Pl} \sim 10^{-33}$ cm.
However, all we know from experimental tests of gravity is its force
at the present time on macroscopic distances, that is in the range
10$^{-2}$ cm -- 10$^{28}$ cm.

Loopholes have been found in recent years. For example, in models with
extra dimensions~\cite{ADD}-\cite{rs} the true fundamental gravity
scale can be as low as a few TeV, and the large Planck mass is then
merely an effective long-distance 4-dimensional parameter. For a
recent review see~\cite{anton-rev}.  In these scenarios, gravity
becomes phenomenologically interesting for high energy physics and we
may observe and study quantum gravity phenomena at future colliders.

Another suggestion to explain the electroweak-gravitational
hierarchy in a natural way in the 4 dimensional world was
recently put forward in Ref.~\cite{notari}. It is
assumed that there exists a scalar field $\Phi$ with
nonminimal coupling to the curvature, $RV(\Phi)$. The 
initial value of the function $V(\Phi)$ is supposed to be  
in electroweak scale, i.e. about (TeV)$^2$, while due to
dynamical evolution of $\Phi(t)$ it may reach the
asymptotic Planck value, $V(\Phi_\infty) = M_{Pl}^2 $.

While TeV-gravity is a fascinating possibility from the point of view
of particle phenomenology, its cosmology may be problematic. In fact
we can reasonably expect that in such models the maximum temperature
of the universe never exceeded a few TeV, since the concept of
spacetime itself exists only for temperatures below the fundamental
gravity scale. In fact we often find the reheating temperature after
inflation to be significantly lower.  Consequently, ordinary cosmology
commenced at temperatures so low that electroweak symmetry breaking
$M_{EWSB} \sim 300$ GeV never took place in the early universe.  Since
at the moment we have no reliable information about the universe
before it was 1 s old, i.e. before primordial nucleosynthesis, there
are no direct contradictions with the assumption of TeV
gravity. However, baryogenesis is quite difficult in these models,
because to this end a mechanism working at relatively low energies is
needed and presently we do not know anything suitable in the context
of the Standard Model.  In particular, violation of baryon number
($B$) conservation below the electroweak phase transition is
completely negligible in the standard theory and this seemingly
forbids any realistic baryogenesis scenario in the case of low scale
gravity.

On the contrary, a low fundamental scale of gravity opens a new
possibility for TeV scale baryogenesis, because the gravitational
interaction itself can naturally break B-conservation.  In this paper
we consider gravitational decays of heavy particles as a mechanism for
low-temperature baryogenesis.  The details of the heavy particle
decays are irrelevant.  Instead, the key feature is that the decays
are mediated by virtual black holes (BHs), which (according to common
belief) can decay/evaporate with violation of global $U(1)$-quantum
numbers including baryonic charge.  

The possibility that BH evaporation could create the matter-antimatter
asymmetry of the universe was suggested in Ref.~\cite{bh-bg} and
considered in detail in Refs. \cite{ad-bg} and \cite{ad-bg2}. The
scenario was criticized in Ref. \cite{ttwz} on the basis that BH
evaporation produces a thermal equilibrium state; yet, in the absence
of CPT violation, a departure from thermal equilibrium is needed in
order to produce an excess of particles over anti-particles. In
response to this criticism, although the particle emission due to
Hawking radiation at the BH horizon is indeed thermal, the equilibrium
distribution is distorted after particle propagation in the
gravitational field of the parent BH~\cite{page} and their mutual
interactions~\cite{ad-bg}.  In any case, in this paper we consider
decay rather than BH evaporation.  We will not deal with thermal
Hawking radiation \cite{hawking}, a semiclassical process that can be
realized only for ``large'' BHs.  Instead, the decays considered here
are essentially quantum gravity phenomena, with a small number of
final particles not emitted with a thermal spectrum.  Other criticisms
may arise if, believing in the information preserving BHs picture, one
were to argue that baryon number is not violated. However, a rigorous
proof is lacking and, on the contrary, very reasonable arguments
suggest that global quantum numbers are not be conserved~\cite{ssa}.

The basic idea of the baryogenesis scenarios considered here is that
TeV mass particles (the mass of the fundamental gravity scale) decay
gravitationally via intermediate BHs; these decays violate
baryon number.  The essential ingredients of these baryogenesis
scenarios are the three standard ``Sakharov conditions''~\cite{sakharov}.  
We stress that these conditions are {\it
easier} to satisfy with a low fundamental scale of gravity.  The first
criterion, baryon number violation, is mediated by virtual BHs which
can violate global quantum numbers; such gravitational effects are
inversely related to the effective Planck mass and hence are stronger
for low fundamental gravity scales.  The second criterion,
CP-violation, which is negligible at high temperatures in the Minimal
Standard Model, may be much larger in TeV gravity models.  First, the
effective temperatures can be quite small, about a few hundred MeV,
and second, we consider time variation of the quark masses and their
mixing angles.  The third criterion, deviation from thermal
equilibrium, which is normally negligible at electroweak energies,
might be amplified by a much faster Hubble rate, which in turn is
enhanced by a very small Planck mass. These features may lead to very
efficient baryogenesis at relatively low energies.

We would like to stress that we do not introduce any new hypothesis
for our baryogenesis scenario. All they are considered in the literature
and we make proper references to them. For example there are two 
possible ways for realization of fundamental gravity scale in TeV
range: either higher dimensions or time variation of the gravitational
coupling constant $G_N(t)$ or what is the same the Planck mass, 
$M_{Pl} (t)$. In the first case the proton decay and
neutron-antineutron oscillations challenge the hypothesis and 
special efforts should be made to avoid contradiction with
experiment at zero temperature. However, at TeV temperatures the
processes with baryon non-conservation could be easily 
unsuppressed. Time variation of $G_N$ does not encounter these
problem because now we have normal Planck scale gravity,
while it was in TeV range in the early universe and baryon
nonconservation was facilitated.

A larger, than standard model, CP-violation can be introduced by
time variation of the Yukawa coupling constants of quarks with the
Higgs boson, which is
also considered in the literature. One may object to this additional
assumption, on the ground that it is unnatural to have variation
of both $M_{Pl}$ and quark masses. However, it may be just 
opposite: if one mass varies with time, the other may vary as well.
It is natural to expect that all the masses were in TeV range at
the early time. This is the assumption that we have done. After that,
as one can easily see, the standard scenario of baryogenesis 
in heavy particle decays very well operates at TeV energies.

The content of the paper is as follows.  We briefly review TeV-gravity
models in Sec.~\ref{s-models} and the related early cosmology in
Sec.~\ref{s-cosmo}.  In Sec.~\ref{s-baryo} we present our baryogenesis
model and give an estimate of the resultant matter-antimatter
asymmetry.  We conclude in Sec.~\ref{s-conclusion}.


\section{TeV-gravity models \label{s-models}}

There has been a great deal of interest recently in models with a low
scale for gravity, especially since they may provide a resolution to the
perplexing hierarchy problem in particle physics. Two possibilities
have been discussed for TeV-scale gravity: 1) large extra dimensions
and 2) time-varying Planck mass. We briefly review these ideas and their
experimental consequences.

{\it Large Extra Dimensions:} 
In 1998 Antoniadis, Arkani-Hamed, Dimopoulos and Dvali proposed a
``geometric'' solution to the hierarchy problem of high energy
physics, where the observed weakness of gravity (at long distances)
would be related to the presence of large compact extra
dimensions \cite{ADD,AADD}.  Motivated by string theory, the
observable universe would be a 4-dimensional brane embedded in a
(4+$n$)-dimensional bulk, with the Standard Model particles confined
to the brane, while gravity is allowed to propagate throughout the
bulk.  In such scenarios, the Planck mass $M_{Pl}$ becomes an
effective long-distance 4-dimensional parameter and the relation with
the fundamental gravity scale $M_{\ast}$ is given by
\begin{eqnarray}
M_{Pl}^2\sim M_{\ast}^{2+n}R^n \; ,
\end{eqnarray}
where $R$ is the size of the extra dimensions.
If these extra dimensions are ``large'', i.e.
$R \gg M_{Pl}^{-1} \sim 10^{-33}$ cm, then the fundamental
gravity scale can be as low as a few TeV and therefore
of the same order of magnitude as $M_{EW}$.
If we assume $M_{\ast} \sim 1$ TeV, we find:
\begin{eqnarray}\label{size}
R \sim 10^{({30}/{n})-17} \: \textrm{cm} \; .
\end{eqnarray}
In this approach, however, the hierarchy problem is not 
really solved but shifted instead from the hierarchy in energies
to a hierarchy in the size of the extra dimensions which are much
larger than 1/TeV $\sim 10^{-17}$ cm but much smaller than the
4-dimensional universe size. 

The case $n = 1$ is excluded because from Eq. (\ref{size})
we would obtain $R \sim 10^{13}$ cm and therefore strong
deviations from  Newtonian gravity at solar system
distances would result. For $n \ge 2$, $R \lesssim 100$ $\mu$m
and nowadays we have no experimental evidence
against a modification of gravitational forces
in such a regime \cite{r2}. Interesting variations
of these models can lower the fundamental gravity
scale with the use of non-compact extra dimensions
\cite{rs}.

If gravitational interactions become strong
at the TeV scale, quantum gravity phenomena are
in the accessible range of future experiments
in high energy physics. In particular, there is
a fascinating possibility that hadron
colliders (such as LHC) will be BH factories
(for a review, see e.g. Ref.~\cite{colliders}, criticisms can be
found in Refs.~\cite{mbv,adams}).
From the classical point of view, we expect
BH production in collision
of two particles with center of mass energy
$\sqrt{s}$, if these particles approach each other 
so closely that they happen to be inside the event 
horizon of a BH with mass $M_{BH}\approx\sqrt{s}$.
Semiclassical arguments, valid for
$M_{BH} \gg M_{\ast}$, predict the BH production
cross-section
\begin{eqnarray}
\sigma \approx \pi R^2_{BH}(M_{BH}) \; ,
\end{eqnarray}
where $R_{BH}(M_{BH})$ is the horizon radius
of a BH of mass $M_{BH}$. 

{\it Time-varying Planck mass:} An alternative origin of a fundamental
TeV scale for gravity involves a time-varying Planck mass.  The idea
that the value of the Planck mass has evolved with time, and was much
lower in the early universe, goes back to Dirac and his ``large number
hypothesis'' \cite{dirac}.  This idea was then developed by other
authors as a complete field theory of gravitation and culminated in
the Brans-Dicke theory \cite{bd} and in more general scalar-tensor
theories of gravity \cite{damour}. These models have been extensively
studied in the literature, but only recently \cite{notari} has it been
stressed that they are capable of solving the hierarchy problem.  In
\cite{notari} the authors take
\be
V(\Phi) \sim M^2_* f(\Phi) \; ,
\ee
where $M_* \sim M_{EW}$ is the only fundamental scale of the theory
and $f(\Phi)$ a dimensionless function of $\Phi$. The huge gap between
$M_{Pl}$ and $M_{EW}$ we observe today is explained with a temporal
evolution of the scalar field $\Phi(t)$ in the 4-dimensional
spacetime. As a modification of the the model of Ref.~\cite{notari}
we can consider, for example, an exponential potential
\be
V(\Phi) = V_0 \exp [W(\Phi)]
\label{v-of-phi}
\ee
with e.g. $W = (\Phi/\mu)^2 - \lambda (\Phi/\mu)^4$. 
A reasonably small $\lambda \sim 10^{-2}$ could 
ensure the required hierarchy of 16 orders of magnitude between the 
Planck and electroweak scales. 
We plan to present elsewhere a detailed study of the evolution of $\Phi$ 
and the features of the corresponding cosmology.  

If the Planck mass depends on the value of a scalar field $\Phi$ and
today has its usual large value with $M_{Pl} \gg M_{EW}$, then
gravitational interactions should be negligible in particle physics
today, as in the standard theory. In particular, the next generation
of colliders will not be able to produce BHs.  Nevertheless, in the
early universe, when $\Phi$ has not yet evolved to its present value,
non-negligible quantum gravity effects might be effective.
Baryogenesis, in particular, could take place as is
described in the present paper.


\section{Early universe in theories with low scale gravity\label{s-cosmo}}

According to the standard hot Big Bang model, which is described by
the Friedmann equation, as we look backwards in time, the universe was
hotter and hotter. According to common belief, such equations,
obtained from classical general relativity, break down when we reach
the temperature $T \sim M_{Pl}$ and curvature $R\sim M_{Pl}^2$, at
which point quantum gravity phenomena become important: it is
reasonable to expect that the universe has never exceeded these values
of curvature and temperature and that the initial singularity in the
Robertson-Walker metric is a drawback of the classical
theory. Supplementing the Big Bang, in order to resolve difficulties
such as the horizon and flatness problems, the inflation paradigm
\cite{inflaton} was introduced.  Inflation requires a superluminal
expansion rate of the very early universe followed by a period of
reheating. All the relics we do not want to be abundant in the
present-day universe (such as superheavy objects capable of
overclosing the universe) must be produced before inflation, so they
can be strongly diluted. On the other hand, events which must have
left traces (such as the baryogenesis) must take place after
inflation.

If the true fundamental gravity scale is in the TeV range, the maximum
conceivable temperature of the universe is of the same order of
magnitude and the reheating temperature is probably too low to allow
for the electroweak phase transition.  Early universe cosmology is
deeply modified and the generation of the matter-antimatter asymmetry
becomes a real challenge.  Popular scenarios, operating at the GUT
scale, $M_{GUT} \sim 10^{16}$ GeV, or at the electroweak symmetry
breaking with $T \sim $ TeV, cannot work. New mechanisms, efficient
at lower energies, are needed.  We note that chain inflation
\cite{fls} with the QCD axion also leads to a low reheat temperature,
$T \sim 10$ MeV.

As for models with large extra dimensions, the situation is even
worse, because the reheating temperature is usually expected to be well
below $M_*$.  In fact, at high temperatures a copious production of
gravitons into the bulk took place; if we require the cosmological
expansion rate compatible with the observations of primordial light
elements created when the temperature of the universe was $T_{bbn}
\sim 1$ MeV, we obtain a maximum temperature (usually assumed as upper
bound for the reheating temperature after inflation) \cite{add2}
\be
T_{max} \lesssim M_* 
\Big(\frac{T_{bbn}}{M_{Pl}}\Big)^{1/(n+2)} \; .
\ee
For $n = 2$ we obtain $T_{max} \lesssim 10$ MeV, whereas $n = 7$ leads
to $T_{max} \lesssim 10$ GeV.  At such low temperatures standard
scenarios of baryogenesis are impossible.

Constraints on  the time variation of the Planck mass can be
derived from different cosmological and astrophysical considerations
(for a review, see e.g. Ref. \cite{uzan}). The most stringent bound
comes from the big bang nucleosynthesis.  The analysis of light
elements production requires that when the universe temperature was
around 1 MeV the Planck mass had to be essentially frozen at its
present value: the allowed deviation from the value that we measure 
today should be less than 5\% \cite{bambi}.

In this paper we will investigate baryogenesis with TeV gravity scales
and low reheating temperature in a model independent way, rather than
distinguishing between models with extra dimensions from models with a
time variable Planck mass.  The essential physics is the same.
General features of baryogenesis with a low gravity scale, based on
enumeration of possible non-renormalizable B-violating operators in an
effective low energy Lagrangian, are described in Ref. \cite{benakli}.
We note that some alternative ideas for baryogenesis at extremely low
temperatures have been considered earlier (see e.g. \cite{banksdine}).
A different picture is considered in Ref. \cite{baby-branes}, where
an effective baryon number violation on our brane could result from
baryon evaporation into ``baby branes'' or from baryon exchange in
brane collisions, so that the higher dimensional spacetime remains
baryon symmetric and the matter dominated universe is reduced to
a peculiar feature of our brane.


\section{Mechanism of baryogenesis \label{s-baryo}}

In order to generate a cosmological matter-antimatter asymmetry
in an initially symmetric universe, usual
baryogenesis scenarios assume CPT invariance and
require the so-called ``Sakharov conditions''
\cite{sakharov}:
\begin{enumerate}
\item baryon number non-conservation,
\item violation of C (charge conjugation) and CP (charge conjugation
combined with parity) symmetries,
\item deviation from thermal equilibrium.
\end{enumerate} 
For a discussion, see e.g. \cite{b-rev,dz,kt}.

Here we consider a possible baryogenesis mechanism in models with
a fundamental gravity scale $M_*$ in the TeV range.  In particular
we discuss the original scenario of out-of-equilibrium decay of
heavy particles, $X$, appropriately modified to the case of TeV scale
gravity. In this section we briefly describe the main features of the
scenario and emphasize the advantages of TeV scale gravity in
satisfying the ``Sakharov conditions'' required for any baryogenesis
model.  We will try to be as close to the Minimal Standard Model in
particle physics as possible, though we do not reject a possible
supersymmetric extension which may make baryogenesis more
efficient. We will present also a more detailed realization
of the three conditions in a concrete model.

1. {\it Baryon number violation:}
In this paper we use gravitational effects that violate baryon
number and we focus on the role of baryon-violating processes
mediated by BHs.
Since such gravitational effects are inversely proportional to a 
power of the
effective Planck mass, a smaller fundamental gravity scale leads to
more effective baryon violation.  Thus a strong non-conservation of
baryonic charge is a generic feature of TeV gravity models.  In fact,
care should be taken to avoid too strong nonconservation of baryons to
keep protons reasonably stable. On the other hand, this feature of 
an enhanced baryon number
violation is favorable for cosmological baryogenesis.

2. {\it CP violation:}
CP-nonconservation in the Minimal Standard Model is known to be
very weak. At high temperatures it is proportional to 
\be \label{cp-suppression}
\epsilon_{CP} \approx
(m_t^2-m_c^2) (m_t^2-m_u^2) 
(m_c^2-m_u^2) (m_b^2-m_s^2) 
(m_b^2-m_d^2) (m_s^2-m_d^2) 
J_{PC} / T^{12}
\ee
where $J_{CP}$ is the Jarlskog invariant
\be
J_{CP} = \cos\theta_{12} \cos\theta_{23} \cos^2\theta_{13}
\sin\theta_{12} \sin\theta_{23} \sin\theta_{13}
\sin\delta_{CP} \approx 3 \cdot 10^{-5} \; .
\label{jarl}
\ee
Here $\theta_{ij}$ are mixing angles
between different generations and $\delta_{CP}$
is the CP odd phase in the mass matrix.
For $T\sim 100$ GeV,  $\epsilon_{CP} \approx  10^{-19}$. 
Such a small magnitude surely demands some modification of
the standard mechanism of CP violation to allow for successful
baryogenesis.

Enhanced CP violation is possible assuming time
dependent quark masses and mixings.  Large CP violation may arise
if quark masses were in the 100 GeV -- TeV range in the 
early universe, with the mass differences of the same order of
magnitude as the values of the masses. It is natural to expect
that simultaneously with the masses, the mixing angles between quarks
also changed and might possibly be of the order unity in the early
universe, because both mixings and masses are determined by 
diagonalization of the same mass matrix which has different entries
in the early universe and today. Since by assumption the quark masses
were of the same magnitude in the early universe, all the mixings
should be also of the same magnitude and quite probably close to unity.

On the other hand, if the temperature after inflation were
much smaller than 100 GeV, $\epsilon_{CP}$ in Eq. (\ref{cp-suppression}) 
might not be so strongly suppressed. For example the reheating temperature
in the MeV range would lead to CP-odd effects of the same order
of magnitude as those observed in K or B mesons decays.

3. {\it Deviation from thermal equilibrium:} Here we focus on
out-of-equilibrium decays of TeV scale particles as responsible for
the generation of the baryon asymmetry.  A sufficient cosmological
abundance of such
particles is not hard to imagine, they may e.g. be created during
reheating after inflation as described further below.

The deviation from thermal equilibrium of non-relativistic
decaying particles at a temperature $T$ is much larger in TeV 
gravity than in the usual Planck scale one. Indeed, the deviation 
is determined by the ratio of the universe's expansion rate $H$ to the
reaction rate $\Gamma \sim g^2 m_X/2\pi$, where $g$ is the coupling 
constant of $X$-particles to lighter decay products. 
Normally $g^2 \sim 0.1$. Hence, for example, in the standard 
cosmology where $H \sim \sqrt{\rho}/M_{Pl}$, 
with $\rho \sim T^4$ being the cosmological
energy density, the parameter describing deviation from 
equilibrium at $T\approx m_X$ is
\be
\delta_{neq} \equiv \frac{H}{\Gamma}
\sim \frac{10^2\,m_X}{ M_{Pl}} \; .
\label{delta-neq}
\ee
One can check that the magnitude of the baryon asymmetry generated in 
heavy particle decays is proportional to $\delta_{neq}$, see e.g. 
Ref.~\cite{ad-cp}.
In the case of electroweak masses and with $M_{Pl} \sim 10^{19}$ GeV, 
$\delta_{neq} \sim 10^{-15}$ and is negligibly small. On the other hand,
if $M_{Pl}$ depends on time and was a few TeV
in the early universe,
$\delta_{neq}$ might be easily of the order of unity.  
Thus a low fundamental gravity scale leads to
out-of-equilibrium decays at much lower temperatures.

As for the braneworlds models, the situation is a little more
subtle, since the related cosmology can be quite different
from the standard one. For example, in the case of one
extra dimension compactified on a circle, the effective
4-dimensional Friedmann equation is \cite{binetruy1,chungfreese}
\be\label{braneworlds}
H^2 \sim \Big(\frac{\rho_{brane}}{48 \pi M_*^3}\Big)^2 
+ \frac{\Lambda_{bulk}}{48 \pi M_*^3} \; ,
\ee
where $\rho_{brane}$ is the total energy density on the
brane and $\Lambda_{bulk}$ a possible bulk cosmological constant.
Because of the compact nature of the extra
dimension, we find that one of the dominant terms of the 00-component 
of the Einstein equations, i.e. the 
the square of the logarithmic derivative of the scale factor
with respect to the extra dimension coordinate, is equal to
$(a'/a)^2 \propto \rho^2_{brane}$. Hence,
in order to recover the standard cosmology at low temperatures
\cite{binetruy2}, $\rho_{brane}$ can be split into the energy density
of ordinary matter $\rho$ and the brane cosmological constant
$\Lambda$ (which can be interpreted as the tension of the brane)
and require
\be
\Lambda_{bulk} = - \frac{\Lambda^2}{48 \pi M_*^3} \qquad
\Lambda = 6 \, \frac{(8\pi)^3 M_*^6}{M_{Pl}^2} \; .
\ee
In this case the 4-dimensional Friedmann equation becomes
\be
H^2 \sim \frac{8 \pi \rho}{3 M_{Pl}^2} \Big(1 +
\frac{\rho}{2 \Lambda}\Big) \; .
\ee
For $M_* \gtrsim 1$ TeV, the universe expansion rate is compatible
with the big bang nucleosynthesis, but it is faster than the
standard one at higher temperature, which is favorable for baryogenesis.

Another mechanism for breaking of thermal equilibrium
commonly considered for electroweak baryogenesis is
due to bubble formation in the first order electroweak 
phase transition. We note an advantage of our model, that such 
a first order electroweak phase transition is not necessary to 
create a large deviation from equilibrium. 
{ The first order electroweak phase transition is
probably excluded by a heavy Higgs boson, if the mass of the 
latter is the same today and in the early universe, and in any
case it can not be useful for baryogenesis in models with a
fundamental gravity scale in TeV range: here the universe must
have arisen at a much lower temperature and the electroweak symmetry
was probably never restored.}

There is also another possible source of the out-of-equilibrium
physics required for successful baryogenesis created by the
bubble collisions at the end of ``chain inflation'' as described 
further below. Out of equilibrium conditions for the bubbles 
are also easier to achieve in TeV gravity versus Planck gravity
because of higher expansion rate in the first case.

\subsection{``First Sakharov Condition'':
Baryon number violating decays \label{ss-b}}

It was argued long ago \cite{zeldo} 
that gravity could induce processes with
nonconservation of baryonic number. In particular,
virtual Planck-mass BHs would induce proton
instability and the expected decay width
estimated by dimensional considerations would be
\be
\label{eq:pdec}
\Gamma_p \sim \frac{m_p^5}{M_{Pl}^4} \; .
\ee
For the normal Planck mass, $M_{Pl} \sim 10^{19}$ GeV, gravitational decays
would be dangerous for very heavy particles only, with masses, say, in
the interval $10^{10}-10^{16}$ GeV; for discussion see
Ref. \cite{ad-heavy}. On the other hand, with a smaller Planck mass in
the TeV range, the baryon number violating processes would become much
more efficient. In fact, Adams {\it et al.} \cite{adams} argued that
experimental limits on the proton lifetime constrain the quantum
gravity scale to be larger than $10^{16}$ GeV.  A possible way to avoid
too short life-time of proton is considered in our paper \cite{noi},
where some other approaches are also discussed and the list of
references is presented.

We have proposed there a conjecture that, just as in classical gravity,
sub-Planck-mass BHs can only exist with zero 
local charge (electric or color) and 
zero angular momentum. In fact, according to classical general 
relativity in $3+1$ dimensions, a charged and rotating point-like
particle with mass $m < M_{Pl}$ cannot form a BH, because its charge
and angular momentum prevent the formation of the event horizon.
Therefore, following
our conjecture which forbids formation of a larger BH
through violation of energy for a small time interval, the
sub-Planck-mass initial states can form 
virtual BHs only with vacuum quantum numbers and baryon
violating decays $\Delta B\neq 0$ should be noticeably suppressed.
If we are interested in the decay of particles with nonzero spin 
and/or electric or color charge (such as all the ``elementary''
particles we know today), the formation of a Schwarzschild 
BH demands production of additional virtual 
particles and hence these processes can proceed only in higher
orders of perturbation theory. Due to this conjecture, proton 
decay is suppressed to the point where it is in agreement with 
experimental bounds. In addition, we predict that neutron-antineutron 
oscillations and anomalous decays of muons, $\tau$-leptons and 
$K$- and $B$-mesons can be quite close to the existing bounds and 
these processes may be found in the near future.

On the other hand, in this paper we are interested in the regime 
where the baryon violating decays are quite rapid. Although the 
rates are suppressed today, it is possible that they were much more 
rapid in the early universe. This is achievable if the mass of quarks
changed with time and reached the true quantum gravity scale in the
TeV range in the early universe: here a point-like particle with
non-vanishing quantum numbers may form a BH and the branching ratio
of B-violating decays may be noticeably enhanced. In particular,
the rate of the B-nonconserving decay $t\rar 2\bar q + l $ 
estimated in accordance with our work~\cite{noi} would be about
$10^{-10}$ in the present day universe. { However, if the mass 
of the $t$-quark changed with time and reached the true quantum gravity
scale in the early universe, the suppression mechanism of 
Ref. \cite{noi} does not work and the branching ratio could be even 
of order one (the initial state is no more below the true Planck mass
and charged and rotating intermediate BHs are therefore allowed)}.  
This makes such decays promising for creation of the cosmological 
baryon asymmetry. These decays may be even more efficient if the 
masses of the weak intermediate bosons also change with time in 
such a way that $W$ and $Z$ would be heavier than the heaviest quark 
(in the early universe this is not necessarily the $t$-quark). In such a 
case the electroweak decays of the type $t\rar Z q$ would be 
forbidden and the usual electroweak decays could proceed only through
exchange of  virtual $W$ and $Z$ bosons. The total decay width would be 
much smaller and the branching ratio of B-nonconserving decays would be
strongly enhanced. For a possible mechanism of time variation of quark 
masses see below Subsection \ref{ss-cp}.

An alternative possibility to time variation of the 
quark masses is an
existence of TeV elementary particles, for example supersymmetric 
partners of the Standard Model particles. The lightest SUSY 
particles are ordinarily stable against decays, because of R-parity,
yet they may be able to produce intermediate BHs and 
consequently decay.  If so, these particles might be responsible 
for baryogenesis but then unfortunately could no longer provide 
the dark matter of the universe.

TeV-particles produced out of thermal equilibrium after inflation 
can therefore decay fast via intermediate BHs and since 
the decay/evaporation of such objects does not conserve any global 
symmetry \cite{ssa}, these processes would violate the baryonic 
quantum number and might create the observed
cosmological baryon asymmetry.

\subsection{``Second Sakharov Condition'': Violation of C and CP \label{ss-cp}}

When the intermediate BH state decays, the emitted particles interact
with each other and, in order to generate a matter-antimatter
asymmetry, C and CP violation must be present in their interactions.
As we have already mentioned above, CP-violation in the standard model
is extremely weak at high temperatures because of a small ratio of the
quark mass difference to the temperature.  One should remember that
the weakness of CP-violation in the standard model is induced by the
smallness of the Yukawa couplings between quarks and the Higgs field
and the amplitude (\ref{cp-suppression}) is the same even in the
unbroken phase when the masses vanish.  One should keep in mind that
for the usual electroweak baryogenesis the temperature should be
above or around 100 GeV because sphalerons are not effective
otherwise.  However, in TeV scale gravity the temperature after
inflation may be very low and the suppression (\ref{cp-suppression})
would be much milder.

A new source of CP-violation suggested recently~\cite{pospel} 
in a simple
extension of the Standard Model may be also useful for the 
mechanism considered here.

Another interesting possibility is time variation of the 
quark masses. The idea was put forward in Ref.~\cite{nir} 
but we suggest here a different realization. We assume that
there exists one more scalar doublet $\chi$,
analogous to the Higgs one, which is strongly coupled to all 
quarks, $g_{\chi}\chi\bar{\psi}\psi$ with $g_{\chi} \sim 1$
(roughly speaking with the same strength as the usual Higgs is
coupled to $t$-quark, or somewhat weaker but not as weak
as the usual Higgs field, $\varphi$, is coupled to the light $u$ and 
$d$ quarks). If $\chi$ acquires vacuum expectation value 
in the TeV range only in the early universe, the quark mass 
differences and their masses could be all about TeV. Such unusual 
behavior can be achieved e.g. if $\chi$ has the potential with 
non-minimal coupling to curvature:
\be
U(\chi) = \lambda |\chi|^4 + \xi |\chi|^2 R \; .
\label{U-of-chi}
\ee
If $\xi R <0$ the vacuum with $\chi=0$ is unstable and
the expectation value of $\chi$ in the true vacuum state 
would be
\be
\langle \chi^2 \rangle = \xi R/2\lambda \; .
\label{chi-vac}
\ee
Since the curvature is proportional to the ratio of the
trace of the matter energy-momentum tensor to the square of the
Planck mass
\be
R = - \frac{8 \pi}{M_{Pl}^2} \, T \; ,
\ee
in the universe today we have
\be
\langle \chi^2 \rangle \approx 
10^{-80} \, \frac{\xi}{\lambda} \, {\rm GeV}^2
\ee
which is negligible for any reasonable value of the ratio
$\xi/\lambda$. On the other hand, in the early universe,
before the radiation dominated epoch, 
$\langle \chi^2 \rangle \sim$ (TeV)$^2$ is certainly
achievable, thanks to the possibility of a much lower
effective Planck mass and a high energy density.
In this picture, the quark mixing angles should be also 
much different from their standard late time values
and the suppression 
due to the Jarlskog determinant (\ref{jarl}) could be absent 
or much milder.

Since $\chi$ is more strongly coupled to quarks one should
take care that this field would not contradict the 
precise electroweak
data. It may be probably achieved if $\chi$ is an order of 
magnitude heavier than the usual Higgs, $\varphi$, and the coupling
to light quarks is not too strong.

\subsection{``Third Sakharov Condition'': Out of equilibrium criterion}

The ``third Sakharov condition'' for baryogenesis is that
the universe be out-of-equilibrium so that any baryon number
that is created is not immediately wiped out by other reactions. 
Inflationary cosmology offers two ways to achieve this criterion:
1) bubble collisions due to a first order phase transition in 
chain inflation and 2) out-of-equilibrium
decays of particles produced during reheating in inflation. We 
discuss both possibilities here.

\subsubsection{Chain Inflation}
In chain inflation, a series of tunneling events takes place, e.g., in
a potential that looks like a tilted cosine \cite{fs, fls}.  The field
tunnels from one high energy minimum of the potential to a lower
energy one, and thence to yet another lower energy minimum until it
reaches zero energy. At each stage the universe inflates by a fraction
of an efolding, adding up to a total of sixty efolds after several
hundred tunneling events.  The phase transitions are first order, with
bubbles of true vacuum nucleating inside the de Sitter space.
Reheating occurs at the last few tunneling events, when bubble
collisions of the final true vacuum take place. While these bubble
collisions are taking place, the universe is out of thermal
equilibrium, so that baryogenesis may take place without allowing 
the reverse reactions 
to destroy the baryons that have been created.  This
mechanism is similar to the bubble collision mechanisms that were
discussed for electroweak baryogenesis (should this transition be
first order).  In addition, the energy difference between minima can
be arbitrary, in this case a fraction of a TeV, if the total height of
the potential is constrained to be below the TeV Planck scale at the
time of inflation, Heavy particles can be produced during reheating, 
and these can subsequently have baryon violating decays (again, out 
of thermal equilibrium).

\subsubsection{Production of heavy particles \label{ss-heavy}}

The period of exponential expansion of the universe, known as
inflation, ends up with ``reheating''. As suggested in
Ref. \cite{linde}, such period provides favorable conditions for
possible baryogenesis.  Many weakly interacting particles can be
abundantly created, even very heavy ones. In addition, their reaction
rates can be slow and life-time sufficiently long, allowing them to
decay out of equilibrium and to give the universe a net baryon number.
In standard rolling models of inflation, the reheating proceeds
through three different stages: first, there is the preheating period,
where the classical inflaton field, $\phi (t)$, oscillates, producing
all the particles it couples to; then, the produced particles (if
heavy and unstable) decay; last, particles produced during the
previous two stages interact with each other and thermalize,
converting the universe from a cold and low-entropy state into a
hot and high-entropy one. Heavy particles can be created during
reheating, even with masses larger than frequency of the inflaton
oscillations.
Specific mechanisms include 
tunneling models of inflation (chain inflation, as described in the 
previous subsection) as well as perturbative, nonperturbative, and 
gravitational particle production in rolling models as discussed below.
\\

{\bf 1. Inflaton decay:} The inflaton $\phi$ could perturbatively
decay into particles if the sum of their masses is smaller than the
effective mass of the inflaton.  As usual it is assumed that the
energy density of the inflaton is smaller than the Planck one.  Hence,
for a TeV mass gravity scale, the height of the potential at the
beginning of inflation must be below the TeV scale.  Most rolling
models of inflation with the usual Planckian gravity require
potentials with $10^{19}$ GeV scale widths and GUT scale heights in
order to produce the appropriate amplitude of density fluctuations
$\delta \rho/\rho \sim 10^{-5}$, so that a TeV Planck mass makes such
models untenable.
On the other hand, such a situation is not formally excluded and 
inflation might start with the potential energy of the inflaton much 
smaller than the effective Planck scale, $\sim$ TeV$^4$.

Another option is hybrid inflation or models with many
scalar fields (e.g. assisted inflation \cite{liddle}), where smaller
mass scales can work.  One of the examples considered in
Ref.~\cite{kal-lin} is a hybrid inflation model with compact extra
dimensions, where inflation (at least its latest stage) occurs only in
our 3-brane and the extra dimensions are already stabilized (though a
previous period of inflation both in the bulk and on our brane was
certainly needed). The potential of the model is
\be
V(\phi,\sigma) = \frac{1}{4V}(M_*^2 - \lambda \sigma^2)^2
+ \frac{1}{2} m^2 \phi^2
+ \frac{\mu^2}{2} \phi^2 \sigma^2 \; .
\ee
The mass of the inflaton field before inflation must be $m \sim
10^{-10}$ eV to obtain density perturbations in agreement with
observations.  After inflation the mass of the inflaton is $\mu
M_*/\sqrt{\lambda} \sim M_*$.  If the mass of the inflaton is that
high, it could decay via B-nonconserving channels, e.g. $\phi \rar
3q\,l$ and $\phi \rar 3\bar q\,\bar l$ with different probabilities,
due to CP-violation, and thus the inflaton decays might generate
cosmological baryon asymmetry.
\\

{\bf 2. Nonperturbative particle production by inflaton:} 
The non-perturbative approach was pioneered in 
papers~\cite{ad-dk,brand-sht} where it was shown that the
production rate that vanishes in the lowest orders of
perturbation theory may
be significant if non-perturbative effects are taken into 
account. In particular, production of particles coupled to
the inflaton field as
\be
h\phi\bar{f}f\,\,\,{\rm and}\,\,\, \lambda\phi^2 b^{\dag}b \; ,
\label{infl-coupl}
\ee
where $f$ and $b$ are respectively fermionic and bosonic fields,
would be only mildly, (as $1/\sqrt{\phi_0}$), suppressed~\cite{ad-dk}, 
despite a large effective mass of the produced particles introduced
by such coupling in the case of large amplitude of the inflaton 
oscillations 
\be
\phi = \phi_0 \cos (m_\phi t + \delta) \; .
\label{phi-of-t}
\ee
The particles
are predominantly produced when $\phi$ passes through 
zero~\cite{ad-dk,riotto-99} and during this (short) time the
mass of the produced particles vanishes. However, if in addition
to the effective mass induced by the coupling (\ref{infl-coupl})
there exists a ``normal'' mass of the created particles
$m_f \bar f f$ or $m_b |b|^2$, the production would be 
strongly, exponentially, suppressed if $m_{f,b}$ is large
in comparison with the characteristic frequency of 
$\phi(t)$~\cite{ad-dk}. If so, only light particles
would be created but they may acquire masses if the electroweak
phase transition took place after the universe (pre/re)-heating.
On the other hand, the effective frequency 
can be large for a large amplitude of the 
formally massless inflaton field, which 
can be realized for the potential $U(\phi) =\lambda \phi^4$.
A natural upper bound  $U(\phi) \leq M_{Pl}^4$ implies
$\phi \leq M_{Pl}/\lambda^{1/4}$. Inflation should stop when
the effective inflaton mass or frequency of oscillations is of the 
order of the Hubble parameter, i.e. 
\be
H^2 = \lambda \phi^4/M_{Pl}^2 \sim \omega^2 = \sqrt{\lambda} \phi^2 \; . 
\label{inf-stop}
\ee
In other words the inflaton starts to oscillate and particle production
begins when $\omega \sim M_{Pl}$. It means that the particles with masses
up the the Planck mass can be created by the inflaton.  

In the case of production of bosons parametric resonance is 
possible~\cite{ad-dk,brand-sht}, which can strongly 
enhance the production rate in the case of wide 
resonance~\cite{kls} and facilitate production of particles with
masses exceeding the mass of inflaton.

All the mechanisms described here allow for production of particles
with masses which may be much larger than the universe temperature
after thermalization. Thus the created massive particles may be
out of equilibrium if their life-time is longer than the Hubble
time.
\\

{\bf 3. Gravitational production:} Gravitational particle 
production in time variable metric~\cite{grav-prod} is efficient only 
when the Hubble parameter, $H$, is not too small in comparison
with the particle mass. Due to conformal flatness of the cosmological
FRW metric, conformally invariant massless fermions and vector bosons 
would not be created~\cite{parker}, but quantum conformal anomaly
eliminates this exclusion and allows for noticeable production of 
even massless gauge bosons~\cite{ad-anom}.

For models with a time dependent Planck mass, gravitational particle
production may be significant.  At the end of inflation, the time
dependent Planck mass and expansion rate are both
$H \sim M_* \sim$ TeV.  The particles produced by external
gravitational field should have energies and/or masses in the same TeV
range. The fraction of the produced heavy particles is model dependent
and, in particular, it depends upon the law of relaxation of the
gravity scale from TeV to the asymptotic Planck value. If the time
dependence of the Newtonian constant is generated by the non-minimal
coupling of a scalar field $\Phi$, to curvature~\cite{notari,ad-nuff},
$\xi R\Phi^2 $, the rate of evolution of $M_* (t) $ is determined by
the potential $U(\Phi)$ or, more generally, $U(\Phi,R)$. The effective
frequency of $\Phi(t)$ may be easily and naturally in the same TeV
range and heavy particles, $X$, are to be produced. Their number
fraction may be noticeable, even close to unity. Moreover, if their
life-time is about $\alpha m_X\sim 0.01 m_X$ which is small in
comparison with the expansion rate, $H\sim M_*$, they may become
dominant, even if their energy density was initially small in
comparison with the energy density of relativistic particles. These
features could lead to very efficient baryogenesis through
B-nonconserving decays of TeV mass particles.

\subsection{Generation of a matter-antimatter asymmetry \label{ss-asy}}

As is argued in the previous subsections, TeV scale gravity looks
favorable for low energy baryogenesis in a slightly modified Minimal
Standard Model of particle physics. We argued that all three 
``Sakharov conditions'' are more easily satisfied with a low 
fundamental scale of gravity.

As shown in Subsection \ref{ss-heavy}, heavy particles $X$ may be
produced after inflation by several reasonable mechanisms. Their
relative number density at production is model dependent but in all
the cases it is not negligibly small (and in fact in some cases we
should take care to not overclose the universe with heavy objects):
$r_X=n_X/n_{tot}\geq 10^{-3}$ is a reasonable guess.  This result
depends upon the concrete scenario of heavy particle, $X$,
production. If they are predominantly created by gravitational field
at the end of inflation their energy density can be
estimated~\cite{grav-prod} as $\rho_X/\rho_{tot} \sim
const\,(m_X/M_{Pl})^2$ with a constant factor of order unity. Thus
this fraction may be close to one. The ratio of the number densities,
$r_X$ would be diluted by the entropy released in the inflaton decay
by the factor $m_X/T_{rh}$, where the (re)heating
temperature $T$ is expected to be in the GeV to MeV
range. 
This simplified estimate follows from the made above statement that
the energy density of the heavy and thermalized particles, 
created by the inflaton decays, are of the same
order of magnitude, i.e. $\rho_X \sim T^4$. Since $\rho_X = n_X m_X$ and 
$n_\gamma \sim T^3$, we find $n_X/n_\gamma \sim T/m_X $.

One should also keep in mind that the life-time of the created
heavy particles is large in comparison with the Hubble time at the
moment of the production and their relative energy density rises in
the course of expansion with respect to the energy density of
relativistic species. This would somewhat increase the effect.

There should also be the entropy suppression factor 
due to annihilation of massive species in thermal equilibrium
universe into photons. If $T_{rh} \sim 1$ GeV this factor is about
0.1. Another small factor comes from the suppression of the CP-odd 
effects in the branching ratio at the
level of $\alpha/\pi\sim 10^{-2}-10^{-3}$ due to necessity of
rescattering in the final state 
{(remember that CP violation arises from the interference 
of loop diagrams with the tree graph)}.
Taken together, these small factors
give the suppression of the baryon asymmetry in the interval
$10^{-6}-10^{-7}$, depending upon the model.  
As we have argued at the beginning of Sec.~\ref{s-baryo}  
the amplitude of CP-violation depends upon the quark mixing angles and
in low temperature baryogenesis it should be about $\epsilon_{CP}(T=0)
\sim 10^{-5}$. If quark masses vary with cosmic time, the mixing angles may
be large in the early universe and $\epsilon_{CP}$ may be of order 
unity both at low and high $T$. 

The estimates presented above are of course very approximate and model
dependent but they show that a model with TeV scale gravity and
time varying quark masses is quite efficient in creating cosmological
baryon asymmetry. Taking all the factors together we expect that
the baryon to photon ratio, $\beta$, can 
be easily equal to the measured value~\cite{WMAP}:
\be
\beta = \frac{n_B}{n_\gamma} = 6\cdot 10^{-10} \; .
\label{beta-obs}
\ee


\section{Conclusion \label{s-conclusion}}

We have considered baryogenesis scenarios in models where the true
quantum gravity scale $M_*$ is in the TeV range. Here, baryon number
can be violated by gravitational decays of TeV-particles, which are
produced out of thermal equilibrium after inflation and quickly decay
through a black hole intermediate state, generating the cosmological
matter-antimatter asymmetry.

We would like to stress that a low reheating temperature which
possibly excludes a period of unbroken electroweak symmetry is not a
problem here but a favorable ingredient of the model, since it
prevents dangerous electroweak sphaleron processes capable of washing
out previously created asymmetry. Moreover, low $T$ allows for a much
larger CP violation from the CKM matrix. In fact, our mechanism
cannot work with the standard Planck mass $M_{Pl} \sim 10^{19}$ GeV
and superheavy particles with masses of the same order of
magnitude. In addition, if $M_*$ is at the level of a few TeV and
heavy elementary particles exist, one can possibly test (and therefore
to reject or to accept) the model in the next generation of hadron
colliders.

If SUSY particles were unstable to decay via these same black holes,
then a possible negative consequence of the model would be the
instability of the lightest supersymmetric particle (LSP), which may
exclude a very nice and testable candidate for cosmological dark
matter.  However, if our conjecture~\cite{noi} is true, the life-time
of the LSP might still be long enough to provide the dark matter,
depending upon the quantum numbers and the mass of the LSP.

Our model requires a minimal extension of the particle content of the
standard model.  The scenario may operate with the standard set of
quarks.  We considered TeV gravity models which may provide a
resolution to the hierarchy problem between electroweak and
gravitational scales, due to either large extra dimensions or
time-varying Planck mass.  One variation we considered requires time
variation of the Planck mass and quark masses created by some new
scalar fields.

The value of the baryon asymmetry is model dependent and cannot be
predicted precisely since it depends upon many unknowns but the same
shortcoming is explicit (or implicit) in all other scenarios of
baryogenesis.

We also mention the possibility that other higher dimensional objects,
such as string balls, p-branes, or black branes may serve as
alternatives to black holes as intermediate states responsible for
baryogenesis, though we have not computed any rates for such
processes.


\section*{Acknowledgments}

We thank F. Urban for useful comments and suggestions.


\end{document}